# Asymptotic Normality of the Maximum Pseudolikelihood Estimator for Fully Visible Boltzmann Machines

Hien D. Nguyen* and Ian A. Wood

September 29, 2014

School of Mathematics and Physics, The University of Queensland, St. Lucia, Brisbane, Australia 4075 (email: h.nguyen7@uq.edu.au).

**Abstract**

Boltzmann machines (BMs) are a class of binary neural networks for which there have been numerous proposed methods of estimation. Recently, it has been shown that in the fully visible case of the BM, the method of maximum pseudolikelihood estimation (MPLE) results in parameter estimates which are consistent in the probabilistic sense. In this article, we investigate the properties of MPLE for the fully visible BMs further, and prove that MPLE also yields an asymptotically normal parameter estimator. These results can be used to construct confidence intervals and to test statistical hypotheses. We support our theoretical results by showing that the estimator behaves as expected in a simulation study.

## 1 Introduction

Boltzmann machines (BMs) were first proposed by [1] as neural network models, but may now be seen as statistical models for the conditional and joint distribution of multivariate binary data. Although the BM may include hidden units, it remains an extremely flexible model even without them. We term the case without hidden units: the fully visible BM (FVBM). The FVBM can be defined as follows.

Let $\boldsymbol{X} = (X_1, X_2, ..., X_d)^T$ be a $d$-variate binary random vector, where $X_j \in \{-1, 1\}$ for $j = 1, ..., d$. If $\boldsymbol{X}$ has probability density function

$$\mathbb{P}(\boldsymbol{X} = \boldsymbol{x}; \boldsymbol{\theta}) = \frac{1}{Z(\boldsymbol{\theta})} \exp\left(\frac{1}{2}\boldsymbol{x}^T \boldsymbol{M} \boldsymbol{x} + \boldsymbol{b}^T \boldsymbol{x}\right), \tag{1}$$

where $\boldsymbol{x}$ is a realization of $\boldsymbol{X}$, $\boldsymbol{M}$ is a $d \times d$ symmetric matrix such that $\operatorname{diag}(\boldsymbol{M}) = \mathbf{0}$, $\boldsymbol{b}$ is a $d \times 1$ vector, and $Z(\boldsymbol{\theta})$ is a normalizing constant, then the model is said to be an FVBM. Here, the superscript $T$ denotes matrix transposition, $\mathbf{0}$ is a zero vector of appropriate dimension, and $\boldsymbol{\theta} = \left(\operatorname{vech}^T(\boldsymbol{M}), \boldsymbol{b}^T\right)^T$ is the parameter vector. The normalizing constant can be written as

$$Z(\boldsymbol{\theta}) = \sum_{\boldsymbol{\xi} \in \{-1,1\}^d} \exp\left(\frac{1}{2}\boldsymbol{\xi}^T \boldsymbol{M} \boldsymbol{\xi} + \boldsymbol{b}^T \boldsymbol{\xi}\right). \tag{2}$$

It is clear that for any high dimensionality $d$, expression (2) becomes intractable. This makes conventional techniques such as maximum likelihood estimation (MLE) of the parameter vector $\boldsymbol{\theta}$ infeasible.

To bypass MLE, there have been numerous developments in algorithms for parameter estimation in BMs. These developments include gradient descent on simulated Kullback-Leibler (KL) divergence [1], alternating minimization of the KL divergence [5], mean-field approximations [13, 18], and minimizing contrastive divergence [9]. To the best of our knowledge, none of the algorithms listed above have been shown to consistently estimate the FVBM parameter vector. In contrast, recent results show that statistical consistency can be proven for estimation by maximum pseudolikelihood [11], score matching [12], and nonlinear least-squares [10].

In this article, we develop upon the work of [11] to show that the maximum pseudolikelihood estimator (MPLE) for the FVBM is not only consistent, but is also asymptotically normal. Upon establishing asymptotic normality, one can construct confidence intervals (CIs), as well as perform hypothesis tests for various parameter vector elements. This article parallels the work of [7] who constructed CIs and hypothesis tests via the bootstrap instead.

We establish asymptotic normality by proving that the regularity conditions of Theorem 5.1 (p. 463) from [14] (as prescribed by [2]) are fulfilled for the FVBM. We then support our proofs with simulations to empirically show that approximate normality is exhibited in relatively small sample sizes.

The article proceeds as follows. Firstly, we introduce the MPLE and outline the consistency result of [11] in Section 2. Secondly, we prove the asymptotic normality of the MPLE for FVBMs in Section 3; we also derive confidence intervals and hypothesis testing procedures in this section. Simulation results are then provided in Section 4, and finally, conclusions are drawn in Section 5.

# 2 Maximum Pseudolikelihood Estimation

Let $\boldsymbol{X}_{(j)} = (X_1, ..., X_{j-1}, X_{j+1}, ..., X_d)^T$ be the vector of elements excluding $X_j$ for $j = 1, ..., d$. In the FVBM, the conditional probabilities of $X_j = x_j$ given

$\boldsymbol{X}_{(j)} = \boldsymbol{x}_{(j)}$ can be written as

$$\begin{aligned} & \mathbb{P}\left(X_j = x_j | \boldsymbol{X}_{(j)} = \boldsymbol{x}_{(j)}; \boldsymbol{m}_j, b_j\right) \\ = & \frac{\exp\left(x_j \boldsymbol{m}_j^T \boldsymbol{x} + b_j x_j\right)}{\exp\left(\boldsymbol{m}_j^T \boldsymbol{x} + b_j\right) + \exp\left(-\boldsymbol{m}_j^T \boldsymbol{x} - b_j\right)} \\ = & \mathbb{P}_j\left(\boldsymbol{x}\right), \end{aligned} \tag{3}$$

where $\boldsymbol{m}_j$ is the $j$th column of $\boldsymbol{M}$.

If we denote a random sample as $\boldsymbol{X}_1, ..., \boldsymbol{X}_n$, where $\boldsymbol{X}_i = (X_{i1}, ..., X_{id})^T$ for $i = 1, ..., n$, then the pseudolikelihood (PL) and log-PL for model (1) can be written as

$$\begin{aligned} & PL\left(\boldsymbol{\theta}; \boldsymbol{x}_1, ..., \boldsymbol{x}_n\right) \\ = & \prod_{i=1}^{n} \prod_{j=1}^{d} \mathbb{P}\left(X_{ij} = x_{ij} | \boldsymbol{X}_{i(j)} = \boldsymbol{x}_{i(j)}; \boldsymbol{m}_j, b_j\right) \\ = & \prod_{i=1}^{n} \prod_{j=1}^{d} \mathbb{P}_j\left(\boldsymbol{x}_i\right), \end{aligned}$$

and

$$\log PL\left(\boldsymbol{\theta}; \boldsymbol{x}_1, ..., \boldsymbol{x}_n\right) = \sum_{i=1}^{n} \sum_{j=1}^{d} \log \mathbb{P}_j\left(\boldsymbol{x}_i\right), \tag{4}$$

respectively. The MPLE can then be defined as

$$\hat{\boldsymbol{\theta}}_n = \arg\max_{\boldsymbol{\theta}} \log PL\left(\boldsymbol{\theta}; \boldsymbol{x}_1, ..., \boldsymbol{x}_n\right).$$

Maximizing the PL as a means of estimation was first suggested in [3] in the context of Markov random fields (MRFs), where the likelihood functions have intractable normalization constants similar to (2). Since then, there has been strong research in establishing the statistical properties of MPLE for MRFs (e.g. [8] and [15]) and vector-variate data (e.g. [2], [6], and [23]). Recent reviews of PL methods can be found in [21], [22], and Ch. 9 of [16].

## 2.1 Consistency of MPLE

In [11], it was shown that the MPLE for the FVBM is consistent. This was achieved by firstly establishing that for each

$$\begin{aligned} \log \mathbb{P}_j\left(\boldsymbol{x}\right) = & \; x_j \boldsymbol{m}_j^T \boldsymbol{x} + b_j x_j \\ & - \log\cosh\left(\boldsymbol{m}_j^T \boldsymbol{x} + b_j\right) - \log 2, \end{aligned}$$

the relevant score functions can be written as

$$\frac{\partial \log \mathbb{P}_j\left(\boldsymbol{x}\right)}{\partial b_j} = x_j - \tanh\left(\boldsymbol{m}_j^T \boldsymbol{x} + b_j\right),$$

and

$$\frac{\partial \log \mathbb{P}_j(\boldsymbol{x})}{\partial \boldsymbol{m}_j} = \boldsymbol{x}\left[x_j - \tanh\left(\boldsymbol{m}_j^T \boldsymbol{x} + b_j\right)\right].$$

The facts

$$\mathbb{E}\left(X_j | \boldsymbol{x}_{(j)}\right) = \tanh\left(\boldsymbol{m}_j^T \boldsymbol{x} + b_j\right),$$

$$\mathbb{E}\left[X_j - \mathbb{E}\left(X_j | \boldsymbol{X}_{(j)}\right)\right] = 0,$$

and

$$\mathbb{E}\left(\boldsymbol{X}\left[X_j - \mathbb{E}\left(X_j | \boldsymbol{X}_{(j)}\right)\right]\right) = \mathbb{E}\left(\boldsymbol{X}\right)\mathbb{E}\left[X_j - \mathbb{E}\left(X_j | \boldsymbol{X}_{(j)}\right)\right],$$

were then used to show

$$\begin{aligned}\mathbb{E}\left(\frac{\partial \log \mathbb{P}_j(\boldsymbol{x})}{\partial b_j}\right) &= \mathbb{E}\left[X_j - \tanh\left(\boldsymbol{m}_j^T \boldsymbol{X} + b_j\right)\right] \\ &= 0,\end{aligned} \tag{5}$$

and

$$\begin{aligned}\mathbb{E}\left(\frac{\partial \log \mathbb{P}_j(\boldsymbol{X})}{\partial \boldsymbol{m}_j}\right) &= \mathbb{E}\left(\boldsymbol{X}\right)\mathbb{E}\left[X_j - \tanh\left(\boldsymbol{m}_j^T \boldsymbol{X} + b_j\right)\right] \\ &= \mathbf{0}.\end{aligned} \tag{6}$$

The Hessian for each $\mathbb{P}_j(\boldsymbol{x})$ was then derived to be

$$\begin{aligned}\boldsymbol{H}_j\left(\boldsymbol{x}; \boldsymbol{m}_j, b_j\right) &= \begin{pmatrix} \frac{\partial^2 \log \mathbb{P}_j(\boldsymbol{x})}{\partial \boldsymbol{m}_j \partial \boldsymbol{m}_j^T} & \frac{\partial^2 \log \mathbb{P}_j(\boldsymbol{x})}{\partial \boldsymbol{m}_j \partial b_j} \\ \frac{\partial^2 \log \mathbb{P}_j(\boldsymbol{x})}{\partial b_j \partial \boldsymbol{m}_j^T} & \frac{\partial^2 \log \mathbb{P}_j(\boldsymbol{x})}{\partial b_j^2} \end{pmatrix} \\ &= -\bar{\boldsymbol{x}}\bar{\boldsymbol{x}}^T \cosh^{-2}\left(\boldsymbol{m}_j^T \boldsymbol{x} + b_j\right),\end{aligned}$$

using the notation $\bar{\boldsymbol{x}} = \left(\boldsymbol{x}^T, 1\right)^T$. Here, we use negative powers to denote multiplicative reciprocation, rather than functional inversion.

From results (5) and (6), and by making the assumption:

**A1** for each $j = 1, ..., d$, the matrix $\mathbb{E}\boldsymbol{H}_j\left(\boldsymbol{X}; \boldsymbol{m}_j, b_j\right)$ is nonsingular negative-definite,

[11] showed that the MPLE is consistent by invoking an M-estimator-like argument (e.g. Theorem 5.7 of [20]). Assumption A1 was required to guarantee the existence of a unique point of convergence.

# 3 Asymptotic Normality

To establish the asymptotic normality of the MPLE for model (1), we invoke the multivariate form of Theorem 2.2 from [2] by proving the veracity of the assumptions of Theorem 5.1 (p. 463) from [14]. These assumptions are as follows.

**B1** The function $\mathbb{P}_j(\boldsymbol{x})$ is identifiable with respect to $\boldsymbol{\theta}$, for each $j$.

**B2** For each $j$, $\mathbb{P}_j(\boldsymbol{x})$ have a common support which does not depend on $\boldsymbol{\theta}$.

**B3** There exists an open subset $\omega$ of $\Omega$ containing the true parameter $\boldsymbol{\theta}^0$ such that for all $\boldsymbol{x} \in \{-1,1\}^d$, the conditional probabilities $\mathbb{P}_j(\boldsymbol{x})$ admits all third derivatives $\left(\partial^3/\partial\theta_{k_1}\partial\theta_{k_2}\partial\theta_{k_3}\right)\mathbb{P}_j(\boldsymbol{x})$ for each $\boldsymbol{\theta} \in \omega$ and $j$.

**B4** The first and second logarithmic derivatives satisfy

$$\mathbb{E}\left(\frac{\partial \log \mathbb{P}_j(\boldsymbol{X})}{\partial \theta_k}\right) = 0, \tag{7}$$

and

$$\begin{aligned} & -\mathbb{E}\left(\frac{\partial^2 \log \mathbb{P}_j(\boldsymbol{X})}{\partial \theta_{k_1}\partial \theta_{k_2}}\right) \\ = \; & \mathbb{E}\left(\frac{\partial \log \mathbb{P}_j(\boldsymbol{X})}{\partial \theta_{k_1}}\frac{\partial \log \mathbb{P}_j(\boldsymbol{X})}{\partial \theta_{k_2}}\right), \end{aligned} \tag{8}$$

for each $j$.

**B5** The matrix

$$\boldsymbol{I}_1(\boldsymbol{\theta}) = -\sum_{j=1}^{d}\mathbb{E}\left(\frac{\partial^2 \log \mathbb{P}_j(\boldsymbol{X})}{\partial \boldsymbol{\theta}\partial \boldsymbol{\theta}^T}\right),$$

is non-singular positive-definite, for each $\boldsymbol{\theta} \in \omega$.

**B6** There exist functions $M_{k_1k_2k_3}(\boldsymbol{x})$, such that

$$\sum_{j=1}^{d}\left|\frac{\partial^3 \log \mathbb{P}_j(\boldsymbol{x})}{\partial \theta_{k_1}\partial \theta_{k_2}\partial \theta_{k_3}}\right| < M_{k_1k_2k_3}(\boldsymbol{x}),$$

for each $\boldsymbol{x} \in \{-1,1\}^d$ and $\boldsymbol{\theta} \in \omega$, where $\mathbb{E}_{\boldsymbol{\theta}^0}\left[M_{k_1k_2k_3}(\boldsymbol{X})\right] < \infty$.

Here, $\theta_k$ is the $k$th component of the vector $\boldsymbol{\theta}$, $\Omega$ is the set of all possible $[d(d+3)/2]$-variate vectors which adhere to the FVBM restrictions on $\boldsymbol{M}$, and $\mathbb{E}_{\boldsymbol{\theta}^0}$ indicates the expectation over the FVBM with parameter vector $\boldsymbol{\theta}^0$, rather than $\boldsymbol{\theta}$.

## 3.1 Main Results

We note that B2 is valid by the definition of model (1). Assumption B3 can be validated by noting that $\mathbb{P}_j(\boldsymbol{x})$ is smooth for any fixed $\boldsymbol{x}$ over the parameter space $\Omega$.

Next, we observe that $\log \mathbb{P}_j(\boldsymbol{x})$ is also smooth with respect to $\boldsymbol{\theta}$ for any fixed $\boldsymbol{x}$, and thus all third logarithmic derivatives exist. Because the third logarithmic derivatives are all continuous with respect to $\boldsymbol{\theta}$, the expectation over the discrete sample space $\boldsymbol{x} \in \{-1,1\}^d$ must be bounded. Hence, B6 follows by noting that $|\mathbb{E}Y| < \infty$ implies $\mathbb{E}|Y| < \infty$, for any random variable $Y$.

Assumption B1 can be validated with the following lemma.

**Lemma 1.** *For each $j = 1, ..., d$, if $\boldsymbol{m}_j \neq \boldsymbol{m}_j^*$ or $b_j \neq b_j^*$, then there exist an $\boldsymbol{x}$ such that $\mathbb{P}_j(\boldsymbol{x}) \neq \mathbb{P}_j^*(\boldsymbol{x})$, where $\mathbb{P}_j^*(\boldsymbol{x}) = \mathbb{P}\left(X_j = x_j | \boldsymbol{X}_{(j)} = \boldsymbol{x}_{(j)}; \boldsymbol{m}_j^*, b_j^*\right)$, $\boldsymbol{m}_j^*$ is a $d \times 1$ vector, and $b_j^*$ is a constant.*

*Proof.* If we assume that $\boldsymbol{m}_j \neq \boldsymbol{m}_j^*$ or $b_j \neq b_j^*$, and $\mathbb{P}_j(\boldsymbol{x}) = \mathbb{P}_j^*(\boldsymbol{x})$, then it follows that

$$\begin{aligned} & \frac{\mathbb{P}\left(X_j = 1 | \boldsymbol{X}_{(j)} = \boldsymbol{x}_{(j)}; \boldsymbol{m}_j, b_j\right)}{\mathbb{P}\left(X_j = -1 | \boldsymbol{X}_{(j)} = \boldsymbol{x}_{(j)}; \boldsymbol{m}_j, b_j\right)} \\ = & \exp\left(2\boldsymbol{m}_j^T \boldsymbol{x} + 2b_j\right) \\ = & \exp\left(2\boldsymbol{m}_j^{*T} \boldsymbol{x} + 2b_j^*\right) \\ = & \frac{\mathbb{P}\left(X_j = 1 | \boldsymbol{X}_{(j)} = \boldsymbol{x}_{(j)}; \boldsymbol{m}_j^*, b_j^*\right)}{\mathbb{P}\left(X_j = -1 | \boldsymbol{X}_{(j)} = \boldsymbol{x}_{(j)}; \boldsymbol{m}_j^*, b_j^*\right)}. \end{aligned} \tag{9}$$

Upon simplification, equation (9) implies that

$$\left(\boldsymbol{m}_j - \boldsymbol{m}_j^*\right)^T \boldsymbol{x} + \left(b_j - b_j^*\right) = 0,$$

which is only true for all $\boldsymbol{x} \in \{-1, 1\}^d$ when $\boldsymbol{m}_j = \boldsymbol{m}_j^*$ and $b_j = b_j^*$. Thus, the result follows by contradiction. □

The validation of B4 can be achieved in two steps. Firstly, condition (7) can be shown using results (5) and (6) from Section 2. Secondly, condition (8) can be proven using the following lemma.

**Lemma 2.** *For arbitrary random variables $Y_1$, $Y_2$ and $Y_3$, with probability density functions $p$ that permit the necessary conditionings, the following statements are true:*

$$\mathbb{E}\left[Y_1 Y_2 Y_3 \mathbb{E}\left(Y_1 | Y_2, Y_3\right)\right] = \mathbb{E}\left[Y_2 Y_3 \mathbb{E}\left(Y_1 | Y_2, Y_3\right)^2\right], \tag{10}$$

$$\mathbb{E}\left[Y_1 Y_2 \mathbb{E}\left(Y_1 | Y_2\right)\right] = \mathbb{E}\left[Y_2 \mathbb{E}\left(Y_1 | Y_2\right)^2\right], \tag{11}$$

*and*

$$\mathbb{E}\left[Y_1 \mathbb{E}\left(Y_1 | Y_2\right)\right] = \mathbb{E}\left[\mathbb{E}\left(Y_1 | Y_2\right)^2\right]. \tag{12}$$

*Proof.* We show result (10) directly by observing that

$$\begin{aligned}
& \mathbb{E}\left[Y_1 Y_2 Y_3 \mathbb{E}\left(Y_1|Y_2, Y_3\right)\right] \\
= & \int\int\int y_1 y_2 y_3 \int y_1^* p\left(y_1^*|y_2, y_3\right) dy_1^* \\
& \times p\left(y_1, y_2, y_3\right) dy_1 dy_2 dy_3 \\
= & \int\int\int y_1 y_2 y_3 \int y_1^* p\left(y_1^*|y_2, y_3\right) dy_1^* \\
& \times p\left(y_1|y_2, y_3\right) p\left(y_2, y_3\right) dy_1 dy_2 dy_3 \\
= & \int\int y_2 y_3 \int y_1 p\left(y_1|y_2, y_3\right) dy_1 \\
& \times \int y_1^* p\left(y_1^*|y_2, y_3\right) dy_1^* p\left(y_2, y_3\right) dy_2 dy_3 \\
= & \int\int y_2 y_3 \left(\int y_1 p\left(y_1|y_2, y_3\right) dy_1\right)^2 p\left(y_2, y_3\right) dy_2 dy_3 \\
= & \mathbb{E}\left[Y_2 Y_3 \mathbb{E}\left(Y_1|Y_2, Y_3\right)^2\right].
\end{aligned}$$

Results (11) and (12) can be shown in a similar manner. □

Using Lemma 2, we can validate (8) by the following lemma.

**Lemma 3.** *For each $j = 1, ..., d$,*

$$\begin{aligned}
& \mathbb{E}\begin{pmatrix} \frac{\partial \log \mathbb{P}_j(\boldsymbol{X})}{\partial \boldsymbol{m}_j} \frac{\partial \log \mathbb{P}_j(\boldsymbol{X})}{\partial \boldsymbol{m}_j^T} & \frac{\partial \log \mathbb{P}_j(\boldsymbol{X})}{\partial \boldsymbol{m}_j} \frac{\partial \log \mathbb{P}_j(\boldsymbol{X})}{\partial b_j} \\ \frac{\partial \log \mathbb{P}_j(\boldsymbol{X})}{\partial b_j} \frac{\partial \log \mathbb{P}_j(\boldsymbol{X})}{\partial \boldsymbol{m}_j^T} & \left(\frac{\partial \log \mathbb{P}_j(\boldsymbol{X})}{\partial b_j}\right)^2 \end{pmatrix} \\
= & -\mathbb{E}\boldsymbol{H}_j\left(\boldsymbol{X}; \boldsymbol{m}_j, b_j\right). \qquad (13)
\end{aligned}$$

*Proof.* Result (13) can be shown by proving the three components:

$$\begin{aligned}
& \mathbb{E}\left(\boldsymbol{X}\boldsymbol{X}^T\left[X_j - \tanh\left(\boldsymbol{m}_j^T\boldsymbol{X} + b_j\right)\right]^2\right) \\
= & \mathbb{E}\left[\boldsymbol{X}\boldsymbol{X}^T \cosh^{-2}\left(\boldsymbol{m}_j^T\boldsymbol{X} + b_j\right)\right], \qquad (14)
\end{aligned}$$

$$\begin{aligned}
& \mathbb{E}\left(\boldsymbol{X}\left[X_j - \tanh\left(\boldsymbol{m}_j^T\boldsymbol{X} + b_j\right)\right]^2\right) \\
= & \mathbb{E}\left[\boldsymbol{X} \cosh^{-2}\left(\boldsymbol{m}_j^T\boldsymbol{X} + b_j\right)\right], \qquad (15)
\end{aligned}$$

and

$$\begin{aligned}
& \mathbb{E}\left(\left[X_j - \tanh\left(\boldsymbol{m}_j^T\boldsymbol{X} + b_j\right)\right]^2\right) \\
= & \mathbb{E}\left[\cosh^{-2}\left(\boldsymbol{m}_j^T\boldsymbol{X} + b_j\right)\right]. \qquad (16)
\end{aligned}$$

We show result (14) directly by observing that

$$\begin{aligned}
& \mathbb{E}\left(\boldsymbol{X}\boldsymbol{X}^T\left[X_j - \tanh\left(\boldsymbol{m}_j^T\boldsymbol{X} + b_j\right)\right]^2\right) \\
= \; & \mathbb{E}\left(\boldsymbol{X}\boldsymbol{X}^T\left[X_j - \mathbb{E}\left(X_j|\boldsymbol{X}_{(j)}\right)\right]^2\right) \\
= \; & \mathbb{E}\left(\boldsymbol{X}\boldsymbol{X}^T X_j^2\right) - 2\mathbb{E}\left(\boldsymbol{X}\boldsymbol{X}^T X_j \mathbb{E}\left(X_j|\boldsymbol{X}_{(j)}\right)\right) \\
& + \mathbb{E}\left(\boldsymbol{X}\boldsymbol{X}^T\left[\mathbb{E}\left(X_j|\boldsymbol{X}_{(j)}\right)^2\right]\right).
\end{aligned}$$

Upon applying result (10) to the final line, we get

$$\mathbb{E}\left(\boldsymbol{X}\boldsymbol{X}^T X_j \mathbb{E}\left(X_j|\boldsymbol{X}_{(j)}\right)\right) = \mathbb{E}\left(\boldsymbol{X}\boldsymbol{X}^T\left[\mathbb{E}\left(X_j|\boldsymbol{X}_{(j)}\right)^2\right]\right),$$

which in conjunction with the fact $X_j^2 = 1$, yields

$$\begin{aligned}
& \mathbb{E}\left(\boldsymbol{X}\boldsymbol{X}^T\right) - \mathbb{E}\left(\boldsymbol{X}\boldsymbol{X}^T\left[\mathbb{E}\left(X_j|\boldsymbol{X}_{(j)}\right)^2\right]\right) \\
= \; & \mathbb{E}\left(\boldsymbol{X}\boldsymbol{X}^T\right) - \mathbb{E}\left(\boldsymbol{X}\boldsymbol{X}^T\left[\tanh^2\left(\boldsymbol{m}_j^T\boldsymbol{X} + b_j\right)\right]\right) \\
= \; & \mathbb{E}\left(\boldsymbol{X}\boldsymbol{X}^T\left[\cosh^{-2}\left(\boldsymbol{m}_j^T\boldsymbol{X} + b_j\right)\right]\right),
\end{aligned}$$

as required. Results (11) and (12) can be used to show results (15) and (16) in a similar manner, respectively. □

Finally, we are left with B5. This is the only assumption that cannot be validated theoretically; although, if one assumes A1, then B5 follows as a consequence. Thus, like [11], we assume A1 and give the following theorem for the asymptotic normality of the MPLE for FVBMs; this assumption guarantees the existence and uniqueness of $\boldsymbol{\theta}^0$.

**Theorem 4.** *If $\boldsymbol{X}_1, ..., \boldsymbol{X}_n$ is an iid sample from density (1) with parameter vector $\boldsymbol{\theta}^0$, and A1 is true, then there exists a sequence $\hat{\boldsymbol{\theta}}_n$, such that $\hat{\boldsymbol{\theta}}_n \xrightarrow{P} \boldsymbol{\theta}^0$ and*

$$\sqrt{n}\left(\hat{\boldsymbol{\theta}}_n - \boldsymbol{\theta}^0\right) \xrightarrow{D} N\left(\boldsymbol{0}, \boldsymbol{I}\left(\boldsymbol{\theta}^0\right)\right), \tag{17}$$

*as $n \to \infty$, where*

$$\boldsymbol{I}_0\left(\boldsymbol{\theta}\right) = \boldsymbol{I}_1^{-1}\left(\boldsymbol{\theta}\right)\boldsymbol{I}_2\left(\boldsymbol{\theta}\right)\boldsymbol{I}_1^{-1}\left(\boldsymbol{\theta}\right),$$

$$\boldsymbol{I}_1\left(\boldsymbol{\theta}\right) = -\sum_{j=1}^{d}\mathbb{E}\left(\frac{\partial^2 \log \mathbb{P}_j\left(\boldsymbol{X}\right)}{\partial\boldsymbol{\theta}\partial\boldsymbol{\theta}^T}\right),$$

*and*

$$\boldsymbol{I}_2\left(\boldsymbol{\theta}\right) = \sum_{j=1}^{d}\sum_{j^*=1}^{d}\mathbb{E}\left[\frac{\partial \log \mathbb{P}_j\left(\boldsymbol{X}\right)}{\partial\boldsymbol{\theta}}\frac{\partial^2 \log \mathbb{P}_{j^*}\left(\boldsymbol{X}\right)}{\partial\boldsymbol{\theta}^T}\right].$$

*Proof.* The result follows from Theorem 2.2 of [2] by validating B1-B4 and B6, and assuming A1. □

Thus, for sufficiently large $n$, the distribution of $\hat{\boldsymbol{\theta}}_n$ is approximately normal, with mean vector $\boldsymbol{0}$ and covariance matrix $n^{-1}\boldsymbol{I}\left(\boldsymbol{\theta}^0\right)$.

## 3.2 Statistical Inference

The normality result can be applied to construct two important inferential tools: asymptotic confidence intervals and hypothesis tests. We proceed to present the Wald-type forms of these constructions, as discussed in Section 9.3.1 of [16].

For the $k$th element of $\boldsymbol{\theta}$, an asymptotic confidence interval with confidence level $CL = (1-\alpha) \times 100$ percent can be given as

$$\left(L_{k,CL}, U_{k,CL}\right) = \left(\hat{\theta}_k \mp \Phi^{-1}\left(1 - \frac{\alpha}{2}\right)\sqrt{\frac{I_{0kk}\left(\boldsymbol{\theta}^0\right)}{n}}\right), \tag{18}$$

where $\alpha \in (0,1)$, $\Phi$ is the standard normal distribution function, and $I_{0k_1k_2}\left(\boldsymbol{\theta}\right)$ is the element in the $k_1$th row and $k_2$th column of the matrix $\boldsymbol{I}_0\left(\boldsymbol{\theta}\right)$.

Now, suppose that we can partition the parameter vector $\boldsymbol{\theta}$ into two components such that $\boldsymbol{\theta} = \left(\boldsymbol{\gamma}^T, \boldsymbol{\delta}^T\right)^T$, where $\boldsymbol{\gamma}$ is an $r$-variate vector of interesting parameter elements. One can test the null hypothesis that $\boldsymbol{\gamma}^0 = \boldsymbol{\gamma}^*$ against the alternative that $\boldsymbol{\gamma}^0 \neq \boldsymbol{\gamma}^*$ by using the Wald-type test statistic

$$W_{\boldsymbol{\gamma}^*} = n\left(\hat{\boldsymbol{\gamma}} - \boldsymbol{\gamma}^*\right)^T \boldsymbol{I}_{0\boldsymbol{\gamma}\boldsymbol{\gamma}}^{-1}\left(\boldsymbol{\theta}^0\right)\left(\hat{\boldsymbol{\gamma}} - \boldsymbol{\gamma}^*\right), \tag{19}$$

where $\hat{\boldsymbol{\theta}}_n = \left(\hat{\boldsymbol{\gamma}}^T, \hat{\boldsymbol{\delta}}^T\right)^T$, and $\boldsymbol{I}_{0\boldsymbol{\gamma}\boldsymbol{\gamma}}^{-1}\left(\boldsymbol{\theta}\right)$ is the sub-matrix of $\boldsymbol{I}_0^{-1}\left(\boldsymbol{\theta}\right)$ with rows and columns corresponding to the parameter elements of $\boldsymbol{\gamma}$. Here, $W_{\boldsymbol{\gamma}^*} \xrightarrow{D} \chi^2_{(r)}$ as $n \to \infty$ under the null hypothesis, where $\chi^2_{(r)}$ is the $\chi^2$-distribution with $r$ degrees of freedom.

We note that in expressions (17), (18), and (19), it is required that $\boldsymbol{I}_0\left(\boldsymbol{\theta}\right)$ be evaluated at the true parameter vector $\boldsymbol{\theta}^0$. As $\boldsymbol{\theta}^0$ is unknown in practice, we can use the estimate

$$\hat{\boldsymbol{I}}_0\left(\hat{\boldsymbol{\theta}}_n\right) = \hat{\boldsymbol{I}}_1^{-1}\left(\hat{\boldsymbol{\theta}}_n\right)\hat{\boldsymbol{I}}_2\left(\hat{\boldsymbol{\theta}}_n\right)\hat{\boldsymbol{I}}_1^{-1}\left(\hat{\boldsymbol{\theta}}_n\right), \tag{20}$$

in place of $\boldsymbol{I}_0\left(\boldsymbol{\theta}^0\right)$, where

$$\hat{\boldsymbol{I}}_1\left(\boldsymbol{\theta}\right) = -\frac{1}{n}\sum_{i=1}^{n}\sum_{j=1}^{d}\frac{\partial^2 \log \mathbb{P}_j\left(\boldsymbol{x}_i\right)}{\partial\boldsymbol{\theta}\partial\boldsymbol{\theta}^T},$$

and

$$\hat{\boldsymbol{I}}_2\left(\boldsymbol{\theta}\right) = \frac{1}{n}\sum_{i=1}^{n}\sum_{j=1}^{d}\sum_{j^*=1}^{d}\frac{\partial \log \mathbb{P}_j\left(\boldsymbol{x}_i\right)}{\partial\boldsymbol{\theta}}\frac{\partial^2 \log \mathbb{P}_{j^*}\left(\boldsymbol{x}_i\right)}{\partial\boldsymbol{\theta}^T}.$$

It can be shown that $\hat{\boldsymbol{I}}_0\left(\hat{\boldsymbol{\theta}}_n\right) \xrightarrow{P} \boldsymbol{I}_0\left(\boldsymbol{\theta}^0\right)$, as $n \to \infty$ (cf. Theorem 7.3 of [4]). Details of alternative CI and test statistic constructions can be found in Section 9.3 of [16]; these include multivariate confidence regions, as well as tests for functional restrictions on the parameter vector.

# 4 Simulations

We now report on a simulation study used to assess the empirical evidence for the theoretical assertions from Section 3. In this study, we generated random samples of sizes $n = 128, 256, 512, 1024$ from a 3-variate FVBM with parameter vector $\boldsymbol{\theta}^0 = \left(\text{vech}^T\left(\boldsymbol{M}^0\right), \boldsymbol{b}^{0T}\right)^T$, where $\boldsymbol{b}^0 = \left(b_1^0, b_2^0, b_3^0\right)^T = \mathbf{0}$, and

$$\boldsymbol{M}^0 = \begin{pmatrix} 0 & m_{12}^0 & m_{13}^0 \\ m_{12}^0 & 0 & m_{23}^0 \\ m_{13}^0 & m_{23}^0 & 0 \end{pmatrix} = \begin{pmatrix} 0 & 0 & 0 \\ 0 & 0 & 0 \\ 0 & 0 & 0 \end{pmatrix}.$$

With the simulated random samples, we computed the MPLE by means of the Nelder-Mead algorithm [17] as implemented through the *optim* function in the $R$ programming environment [19]. The MPLE was then used to construct $CL = 90, 95, 99$ percent CIs for each element of $\boldsymbol{\theta}^0$. We also computed the statistic $W_{\boldsymbol{b}^*}$ for testing the null hypothesis that $\boldsymbol{b}^0 = \boldsymbol{b}^* = \mathbf{0}$, against the alternative that $\boldsymbol{b}^0 \neq \boldsymbol{b}^*$ at the $\alpha = 0.1, 0.05, 0.01$ significance levels.

Using $N = 1000$ repetitions of each construction, we assessed the performance of the CIs by computing the coverage percentage (CP): the percentage of times $\theta_k^0 = 0$ appears in $(L_{k,CL}, U_{k,CL})$ out of the $N$ CIs constructed.

Similarly, using $N$ repetitions of each test statistic, we assessed the performance of $W_{\boldsymbol{b}^*}$ by computing the false positive rate (FPR). The FPR is defined as the proportion of times $W_{\boldsymbol{b}^*} > \chi^2_{(3),\alpha}$ out of the $N$ repetitions, where $\chi^2_{(r),\alpha}$ is such that $\mathbb{P}\left(Y > \chi^2_{(r),\alpha}\right) = \alpha$ if $Y \sim \chi^2_{(r)}$.

By Theorem 4, the CP and FPR should approach $CL$ percent and $\alpha$, respectively, as $n$ becomes large.

## 4.1 Results

We present the CP and FPR for the various confidence and significance levels in Tables 1 and 2, respectively.

Upon inspection, the results from both tables show that for all sample sizes tested, CPs and FPRs are close to the anticipated confidence and significance levels, respectively; thus supporting the result of Theorem 4. The deviation from the expected $CL$ and $\alpha$ values appear to be uniform across all values of $n$. This may be due to two factors: either the number of repetitions $N$ is not sufficiently large for distinguishing between the sample sizes $n$; or the probabilistic convergence to the desired confidence or significance level is slow, and larger sample sizes are needed to observe differences in accuracy due to increasing values of $n$.

# 5 Conclusions

In this article, we have extended upon the results of [11] to show that the MPLE for the FVBM is not only a consistent estimator, but is also asymptotically normal. This theoretical result was then supported by a simulation study which showed that the CIs and hypothesis tests constructed using the asymptotic normality of the MPLE exhibit correct levels of confidence and significance, respectively.

Table 1: CPs (in percentages) from $N = 1000$ repetitions for the elements of $\boldsymbol{\theta}^0$ at $CL = 90, 95, 99$ and $n = 128, 256, 512, 1024$.

| $n$ | $CL$ | CP | | | | | |
|---|---|---|---|---|---|---|---|
| | | $m_{11}^0$ | $m_{12}^0$ | $m_{23}^0$ | $b_1^0$ | $b_2^0$ | $b_3^0$ |
| | 90 | 92.0 | 90.5 | 90.6 | 90.5 | 93.7 | 92.6 |
| 128 | 95 | 95.5 | 93.8 | 94.6 | 94.5 | 95.6 | 96.0 |
| | 99 | 98.9 | 97.8 | 98.8 | 98.8 | 99.5 | 98.6 |
| | 90 | 89.8 | 89.9 | 90.9 | 92.4 | 93.1 | 94.6 |
| 256 | 95 | 93.8 | 93.8 | 95.4 | 95.0 | 96.5 | 97.5 |
| | 99 | 98.6 | 97.7 | 98.7 | 99.2 | 99.1 | 99.6 |
| | 90 | 90.1 | 90.3 | 89.6 | 94.4 | 92.5 | 92.1 |
| 512 | 95 | 93.8 | 94.1 | 94.3 | 97.5 | 95.8 | 95.1 |
| | 99 | 97.9 | 97.9 | 98.0 | 99.5 | 99.3 | 99.2 |
| | 90 | 90.5 | 91.5 | 90.1 | 92.6 | 92.7 | 91.3 |
| 1024 | 95 | 95.0 | 95.3 | 94.1 | 95.8 | 96.0 | 95.1 |
| | 99 | 98.5 | 98.7 | 97.9 | 98.9 | 99.1 | 98.8 |

Table 2: FPRs from $N = 1000$ repetitions for the test statistic $W_{\boldsymbol{b}^*}$ at $\alpha = 0.01, 0.05, 0.10$ and $n = 128, 256, 512, 1024$.

| $n$ | $\alpha$ | | |
|---|---|---|---|
| | 0.01 | 0.05 | 0.10 |
| 128 | 0.020 | 0.069 | 0.115 |
| 256 | 0.012 | 0.059 | 0.099 |
| 512 | 0.010 | 0.056 | 0.109 |
| 1024 | 0.015 | 0.055 | 0.110 |

To the best of our knowledge, research in estimation of BMs has mainly concentrated on numerical properties, as oppose to statistical properties. We hope that this article provokes more interest in exploring the application of BMs as a statistical model such as its application in [7].

We only considered the fully-visible case of the BM in this article because it is the simplest to understand in a statistical context. However, in future work, it would be interesting to extend the results to the standard case where hidden units are present.